\pdfoutput=1

\documentclass[twocolumn]{aastex63}

\usepackage{amssymb,amsmath}

\received{June 1, 2019}
\revised{January 10, 2019}
\accepted{\today}
\submitjournal{AJ}

\shorttitle{Effect of multipole moments in a black hole potential}
\shortauthors{Dubeibe et al.}

\begin{document}

\title{Effect of multipole moments in the weak field limit of a black hole plus halo potential}

\correspondingauthor{Euaggelos E. Zotos}
\email{evzotos@physics.auth.gr}

\author[0000-0002-0793-3255]{Fredy L. Dubeibe}
\affiliation{Facultad de Ciencias Humanas y de la Educaci\'on, \\
Universidad de los Llanos, \\
Villavicencio 500017, Colombia}

\author{Tareq Saeed}
\affiliation{Nonlinear Analysis and Applied Mathematics (NAAM)-Research Group, \\
Department of Mathematics, \\
Faculty of Science, \\
King Abdulaziz University, \\
P.O. Box 80203, Jeddah \\
21589, Saudi Arabia}

\author[0000-0002-1565-4467]{Euaggelos E. Zotos}
\affiliation{Department of Physics, School of Science, \\
Aristotle University of Thessaloniki, \\
GR-541 24, Thessaloniki, Greece}

\begin{abstract}

In this paper, we consider a Newtonian system whose relativistic counterpart describes a superimposed halo with a black hole. Our aim is to determine how the quadrupole and octupole moments affect the nature of the motion of a test particle, moving in the close vicinity of the black hole. The different types of trajectories for the test particle are mainly classified as bounded, collisional, and escaping, by using modern color-coded basin diagrams. Moreover, an additional analysis is carried out for distinguishing between the different types of bounded motion (regular, sticky, and chaotic). Our results strongly indicate that the multipole moments, along with the total orbital energy, highly affect the final state of the test particle, while at the same time the basin geometry of the phase space tends to be highly dominated by collision and escape orbits.

\end{abstract}


\section{Introduction}	
\label{intro}

In Newtonian gravity, the exterior gravitational potential of a given source may be written as a series expansion of the inverse distance function $1/r$, termed the multipole expansion potential (henceforth MEP), which provides a measure of the deviation of the body from spherical symmetry. In particular, the multipole moments of Earth have been studied using the data supplied by the GOCE, LAGEOS, and GRACE satellites aiming to map its gravitational field \citep{DFHMP03,CPP12,V99}. In practice, the deviations of planets and stars from spherical symmetry are almost negligible and hence the higher-order multipole contribution is small, however, this is not true for compact objects mainly due to fast rotations \citep{LP99}.

General relativity is the appropriate framework when compact objects are involved, nevertheless, due to the nonlinearity of the Einstein's field equations, the determination of the multipole structure is not a straightforward procedure as the MEP of Newtonian mechanics \citep{FHP89,HP90}. The multipole moments are of fundamental importance in the context of general relativity, where it is a well-established result that the space-time is fully determined by the multipolar structure of a source \citep{BS80,SS98,SHD10}, or in other words, the scalar multipoles are used to identify the spacetime, in like manner that the Newtonian multipole moments describe the Newtonian gravitational field \citep{SP05}. Moreover, in the Newtonian limit, the set of relativistic multipole moments of mass reduce to the multipole moments in Newtonian theory \citep{Q90}.

Bearing in mind that the multipole moments represent the intrinsic structure of the source, in the context of general relativity several investigations have been carried out aiming to determine their influence on the geodesic motion of test particles \citep[see e.g.,][]{GLM08,RPL11,LWH17,WCJ18}. In \citet{VL96}, the dynamic effect of quadrupolar and octupolar moments describing a superimposed halo with a black hole, within the framework of exact solutions in general relativity, has been investigated. The main conclusion of this work is that the quadrupole term does not introduce chaos by itself into the system, while the octupolar term is an important source in the generation of chaos. Later, the scattering of test particles in presence of core-shell gravitational models introduced to describe the inner regions of elliptical galaxies was considered in \citet{ML00}, finding that there is no detectable chaos when oblate halos are present. On the other hand, two independent teams used specific values of the quadrupolar deformation to analyze the geodesic motion around astrophysical objects with non-isotropic stresses, concluding that chaotic motions for oblate and prolate deformation are possible \citep{GL02,DPS07}. As a general conclusion of all these studies, it can be inferred that breaking the reflection symmetry about the equatorial plane allows the occurrence of chaotic behavior of orbits, being a necessary but non-sufficient condition.

Concerning the Newtonian counterpart of relativistic systems and their respective changes in the dynamic behavior, the limiting cases of the relativistic system associated to exact relativistic core-shell models, have been previously studied, where it is found that the relative extents of chaotic zones in the relativistic cases are significantly larger than in the Newtonian models \citep{VL99}. Also, in \citet{IIY15} it is shown that the Newtonian equations of motion of a black ring provide a nontrivial constant of motion quadratic in momenta, concluding that geodesic chaos is caused by relativistic effects. Then, one may think that there should exist a mechanism underlying classical chaos as a consequence of the correspondence principle, which states that the classical limit of general relativity is Newtonian mechanics. In the relativistic case, as mentioned above, the multipole moments uniquely determine the characteristics of the source, therefore, it should be possible to get some hints about this underlying mechanism if efforts are focused on the intrinsic parameters of the system.

Given the above, in the present paper we study the Newtonian limit of a relativistic system that describes a superimposed halo with a black hole, which contains as free parameters the quadrupolar and octupolar moments. Seeking to reveal the effect of higher-order multipole moments on the existence and stability of the fixed points and the dynamics of a test particle orbiting the source, we perform a thorough and systematic numerical study of this system, which shall be compared with the relativistic results presented in \citet{VL96}. The article is organized as follows: In Section \ref{sol} the exact black hole plus halo solution is presented and the Newtonian potential is derived. The existence and stability of fixed points are discussed in Section \ref{eqpts}. The types and classification of orbits are discussed in Section \ref{clas}, using different planes of representation. Finally, the most important conclusions of this investigation are outlined in Section \ref{conc}.

\section{Exact solution and Newtonian potential}
\label{sol}

The general form of a static, axisymmetric metric in quasi-cylindrical Weyl coordinates can be written as
\begin{equation}
d s^{2} = e^{2 \nu} d t^{2} - e^{-2 \nu}\left[e^{2 \gamma}\left(d z^{2} + d r^{2}\right) + r^{2} d \phi^{2}\right],
\end{equation}
where $\nu$ and $\gamma $ are only functions of $(r, z)$. Under these conditions, Einstein's field equations in vacuum reduce to
\begin{eqnarray}
&&\gamma_{, z} - 2 r \nu_{, r} \nu_{, z} = 0,\label{eq1}\\
&&\gamma_{, r} - r\left(\nu_{, r}^{2} - \nu_{, z}^{2}\right) = 0,\label{eq2}\\
&&\nu_{, r r} + \frac{\nu_{, r}}{r} + \nu_{, z z} = 0,\label{eq3}
\end{eqnarray}
with (\ref{eq3}) the Laplace equation in cylindrical coordinates. A particular solution to this system of equations was derived using a prolate spheroidal coordinate transformation $ (u, v) $ \citep{VL96}. In the new coordinate system, the solution to the Laplace equation can be expressed in terms of the Legendre polynomials $P_{n}(x)$, such that the metric function $\nu$ can be written as the superposition of polynomials
\begin{equation}
\nu(u, v)  =a_{0} Q_{0}(u) + b_{2} P_{2}(u) P_{2}(v) + b_{3} P_{3}(u) P_{3}(v).
\end{equation}
Here, the first term corresponds to the mass monopole, while the second and third terms denote the multipolar structure of the halo. The metric function $\gamma(u,v)$ is then calculated by replacing $\nu(u,v)$ in Eq. (\ref{eq1}) or Eq. (\ref{eq2}). For the sake of completeness, and following the procedure outlined in \citet{VL96}, the explicit metric functions are given as
\begin{eqnarray}
2 \nu &=& \log \left(\frac{u - 1}{u + 1}\right) + \nu_{Q}(u, v) + \nu_{O}(u, v),\label{eqnu}\\
2 \gamma &=& \log \left(\frac{u^{2} - 1}{u^{2} - v^{2}}\right) + \gamma_{Q}(u, v) + \gamma_{O}(u, v)\nonumber\\
&&+ \gamma_{Q O}(u, v),\label{eqgamma}
\end{eqnarray}
with

\begin{eqnarray}
\nu_{Q}&=&\frac{Q}{3}\left(3 u^{2}-1\right)\left(3 v^{2}-1\right),\\
\nu_{O}&=&\frac{\Theta}{5} u v\left(5 u^{2}-3\right)\left(5 v^{2}-3\right),\\
\gamma_{Q}&=&-4 {Q} u\left(1-v^{2}\right)+\left({Q}^{2} / 2\right)
\left(u^2-1\right)\left(v^2-1\right) \nonumber\\&&\times \left[u^2 \left(9
   v^2-1\right)-v^2+1\right],\\
\gamma_{O}&=&\frac{2}{5} \Theta \left[v \left(15 u^2 \left(v^2-1\right)-5
   v^2+9\right)-4\right]
+\frac{3}{100}\Theta^{2}\nonumber\\
&& \left(u^2-1\right) \left(v^2-1\right) \left\{5 \left[5 u^4 \left(25 v^4-14 v^2+1\right)\right.\right.\nonumber\\
&&\left.\left.+u^2 \left(50 v^2-70 v^4-4\right)+5 v^4\right]-20 v^2+7\right\},\\
\gamma_{Q O}&=& \frac{6}{5} Q \Theta u \left(u^2-1\right) v \left(v^2-1\right)
   \left[5 u^2 \left(3 v^2-1\right)+3\right.\nonumber\\
   &&\left.-5 v^2\right].
\end{eqnarray}

\begin{figure*}
\centering
\includegraphics[width=0.7\hsize]{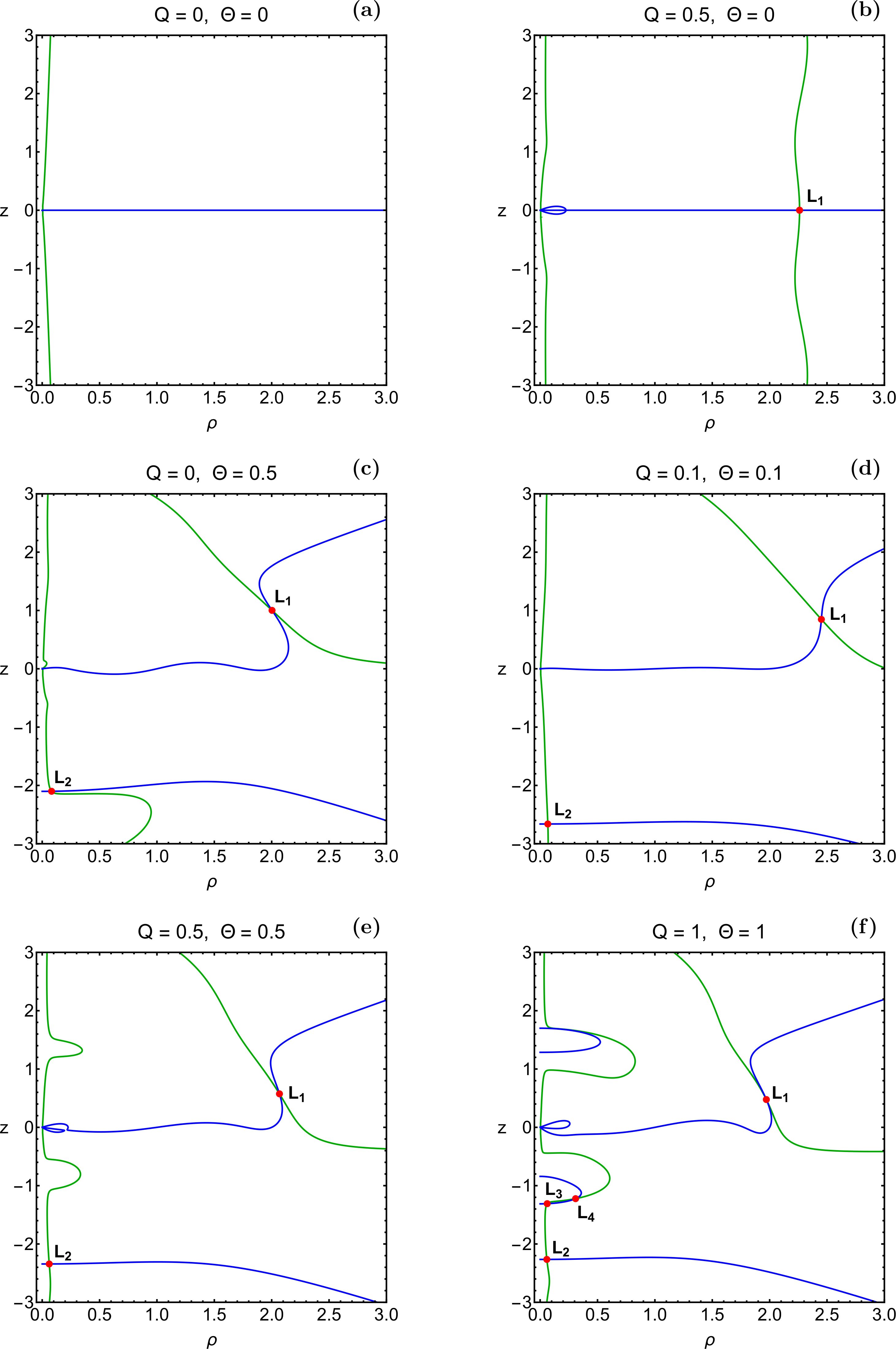}
\caption{Diagrams showing the locations (red dots) of the fixed points through the intersections of the iso-contour lines of the equations $\partial U_{\rm eff}/\partial{\rho} = 0$ (green) and $\partial U_{\rm eff}/\partial{z} = 0$ (blue), for the case where (a): no equilibria exist; (b): 1 point of equilibrium exists; (c-e): 2 points of equilibrium are present; (f): 4 equilibria exist.}
\label{conts}
\end{figure*}

Since we are interested in the physics at the Newtonian regime, and taking into account that in the weak field limit the gravitational potential can be expressed in terms of the metric function as $g_{tt}=1+2\Phi$ \citep[see e.g.,][]{W10}, we expanded the metric function $g_ {tt}=e^{2\nu}$. By using natural units and then transforming to the Euclidean cylindrical coordinates $(\rho,z)$, the Newtonian gravitational potential read as
\begin{equation}
\label{potc}
\Phi \approx - \frac{1}{\sqrt{\rho^2 + z^2}} + \frac{1}{2} \mathcal{Q} f_{\mathcal{Q}}(\rho,z) + \frac{1}{2} \Theta f_{\Theta}(\rho,z),
\end{equation}
where $Q$ and $\Theta$ respectively denote the quadrupole and octupole moments of the source, while
\begin{eqnarray}
f_{\mathcal{Q}}(\rho,z) &=& 2 z^2-\rho^2-\frac{1}{3} \left(\frac{3 z^2}{\rho^2+z^2}-1\right)\left(12 \sqrt{\rho^2+z^2}\right. \nonumber\\
   &&\left.-14 + \frac{4}{\sqrt{\rho^2+z^2}}\right)\\
f_{\Theta}(\rho,z) &=&-\frac{1}{5} z \left(\frac{5 z^2}{\rho^2+z^2}-3\right) \left(25 \sqrt{\rho^2+z^2}-42\right.\nonumber\\
 &&-\left.\frac{4}{\rho^2+z^2}+\frac{26}{\sqrt{\rho^2+z^2}}\right)-3 \rho^2 z+2 z^3
\end{eqnarray}
The time evolution of the system, the conserved quantities, and other properties of the astrophysical system, can be derived from the Lagrangian which in cylindrical coordinates $(\rho, \theta, z)$ can be written as
$
\mathcal{L} = (\dot{\rho}^2 + \rho^2\dot{\theta}^2 + \dot{z}^2)/2 - \Phi(\rho,z),
$
with $\Phi$ given by Eq. (\ref{potc}). The (conserved) momentum conjugate to the cyclic coordinate $\theta$ is $L = \rho^2 \dot{\theta}$ which is associated to the angular momentum about the $z$-axis. Therefore, the Hamiltonian for the 2-dimensional Newtonian system can be written as
$
{\cal{H}} = \left(p_{\rho}^2 + p_{z}^2\right)/2 + U_{\rm eff},
$
where $U_{\rm eff} = L^2/2\rho^2 + \Phi$ is the effective potential, while $(p_\rho, p_z)$ are the conjugated canonical momenta associated to the coordinates $(\rho,z)$, respectively. The corresponding equations of motion in compact form read as
\begin{equation}
\label{eqm}
\dot{\rho} = p_{\rho}, \quad
\dot{z} = p_{z}, \quad
\dot{p_{\rho}} = - \frac{\partial U_{\rm eff}}{\partial \rho}, \quad
\dot{p_{z}} = - \frac{\partial U_{\rm eff}}{\partial z}.
\end{equation}

\section{Equilibrium points}
\label{eqpts}

As indicated in the introduction section, the dynamics of the relativistic system (\ref{eqnu}-\ref{eqgamma}) was previously studied in \citet{VL96}, finding that: a) when $Q = \Theta = 0$ the system is regular, b) when $Q = 0, \Theta \neq 0$, the system exhibits zones of chaotic motion, c) when $Q \neq 0, \Theta = 0$, the system is regular, and d)  when $Q \neq 0, \Theta \neq 0$, the system is chaotic. These results indicate that the presence of an octupole moment of mass is a necessary condition to have geodesic chaos in the system. In view of the above findings, let us start considering the fixed points for the Newtonian limit of this space-time.

The positions of the equilibrium points of the system can be calculated by solving the algebraic system of equations resulting from equating to zero the first-order derivatives of the effective potential
\begin{equation}
\frac{\partial U_{\rm eff}}{\partial \rho} = \frac{\partial U_{\rm eff}}{\partial z} = 0.
\label{pts}
\end{equation}

The solutions to Eq. (\ref{pts}) show that the total number of equilibria is strongly affected by the quadrupole and octupole moments, i.e.
\begin{itemize}
  \item For $Q = \Theta = 0$ there are no fixed points.
  \item For $Q = 0$ and $\Theta \neq 0$, there exist always two equilibrium points.
  \item For $\Theta = 0$ and $Q \neq 0$, there exist always one equilibrium point.
  \item For $\Theta \neq 0$ and $Q \neq 0$ there are either two or four equilibrium points.
\end{itemize}

\begin{figure}
\centering
\includegraphics[width=\hsize]{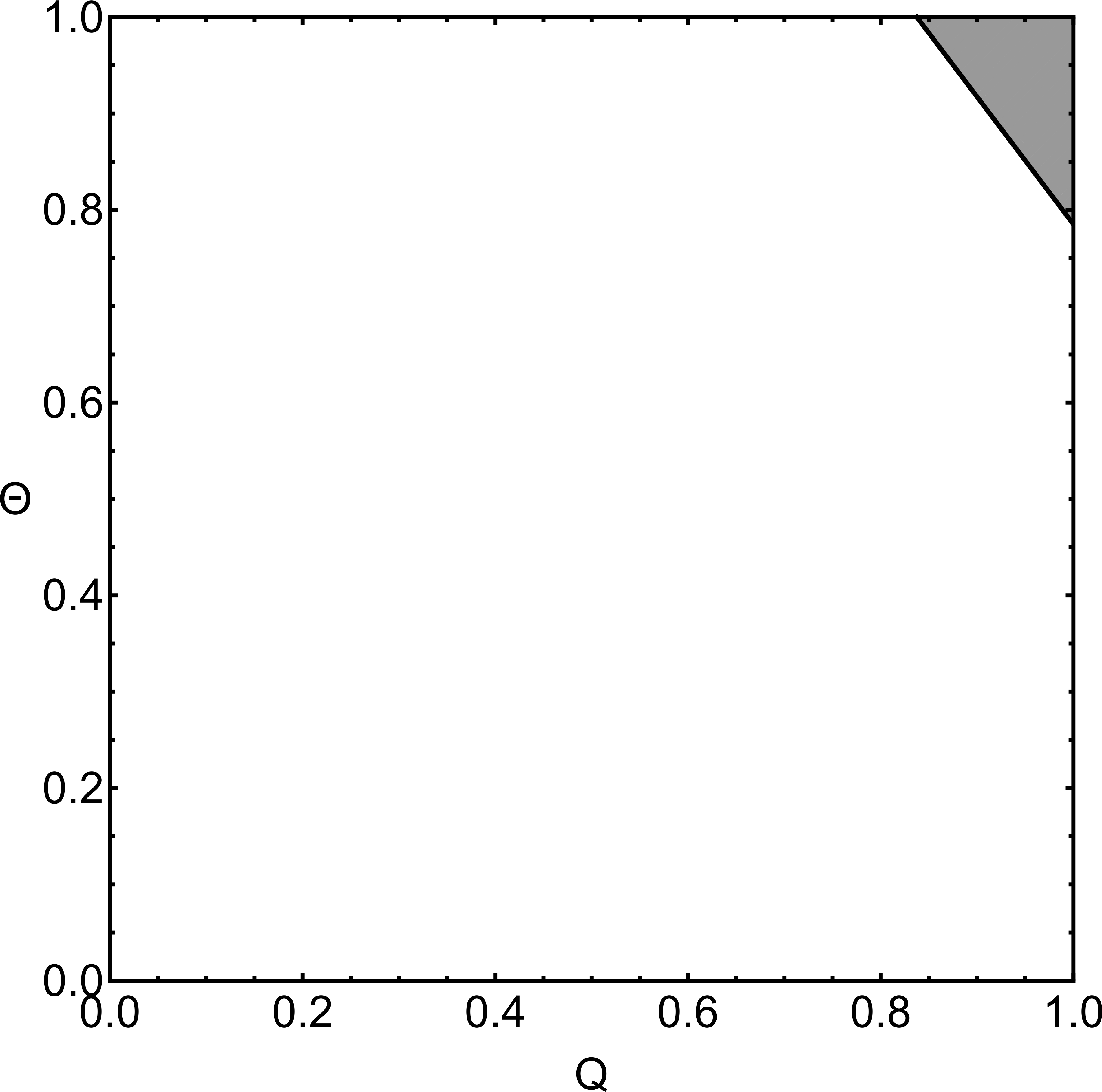}
\caption{The gray-shaded region indicates the set of values $(Q,\Theta)$ for which the libration point $L_3$ is linearly stable.}
\label{regs}
\end{figure}

In Fig.~\ref{conts} we present four characteristic examples corresponding to the four possible cases, regarding the number of equilibrium points. The positions of the equilibria (red dots) are the intersection points of the curves $\partial U_{\rm eff}/\partial{\rho} = 0$ (green) and $\partial U_{\rm eff}/\partial{z} = 0$ (blue). Furthermore, the stability of the equilibrium points can be determined through the standard procedure for linear stability analysis at fixed points. This scheme indicates that when 1 or 2 points of equilibrium exist, they are always linearly unstable. On the other hand, in the case of 4 libration points for relatively high values of the quadrupole and octupole moments, $L_3$ is linearly stable, while the additional three equilibrium points are always linearly unstable. The gray-shaded region in the diagram of Fig.~\ref{regs} shows the set of values of $Q$ and $\Theta$ for which the equilibrium point $L_3$ is linearly stable.

From our previous results on the existence and stability of the fixed points, it can be concluded that the largest number of equilibria and the presence of stable fixed points is only possible when $\Theta \neq 0$, i.e. if the octupole moment of mass exists. Also from Fig.~\ref{conts}, it is observed that the existence of the octupole moment breaks the reflection symmetry along the $z$-axis, due to the presence of odd powers in $z$ for the third term in the potential expansion Eq. (\ref{potc}). This last characteristic of the Newtonian potential is shared with the relativistic counterpart of the system.

\section{Orbit Classification}
\label{clas}

The orbit classification for test particles in the presence of the Newtonian potential Eq. (\ref{potc}), was carried out taking into account that the canonically conjugate quantity to the time is the energy, so Noether's theorem states that the energy is a conserved quantity ${\cal{H}} = E$ and therefore the effective phase space is only three dimensional. The trajectories are classified into three types according to the final fate of the orbit:

\begin{itemize}
\item Bounded orbits, which stay inside the scattering region for $t \rightarrow \infty$.
\item Unbounded orbits, i.e., orbits that escape to infinity for $t < \infty$.
\item Collision orbits, which eventually collide with the central object for $t < \infty$.
\end{itemize}

\begin{figure*}
\centering
\includegraphics[width=0.7\hsize]{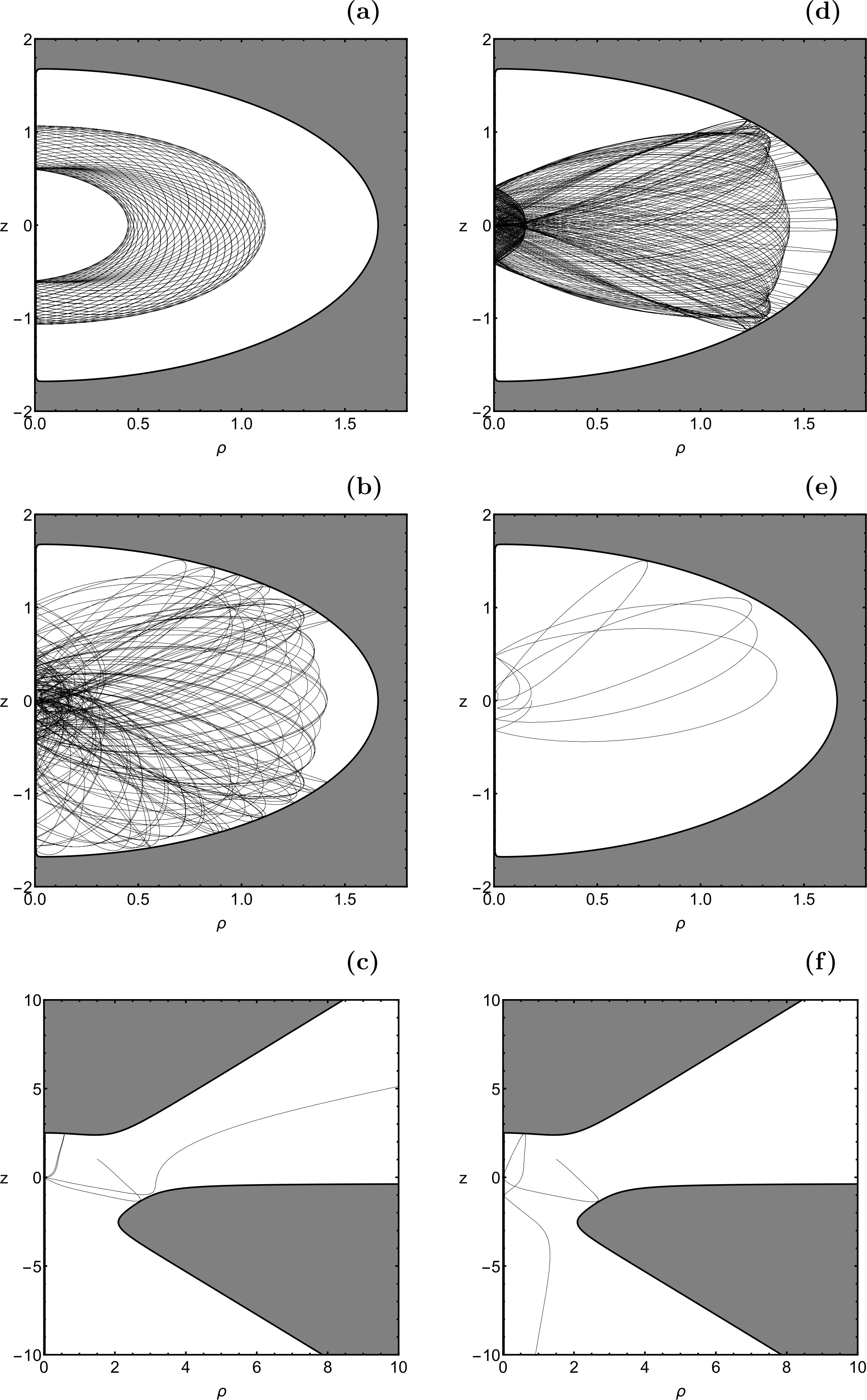}
\caption{Schematic examples of the outcomes of our orbit classification in the configuration space of the system $(\rho, z)$. Each panel shows the type of orbit considered in the analysis: (a): regular, (b): sticky, (c): chaotic, (d): collisional, (e): escaping through channel 1, and (f): escaping through channel 2.}
\label{orbs}
\end{figure*}

\begin{figure*}
\centering
\includegraphics[width=\hsize]{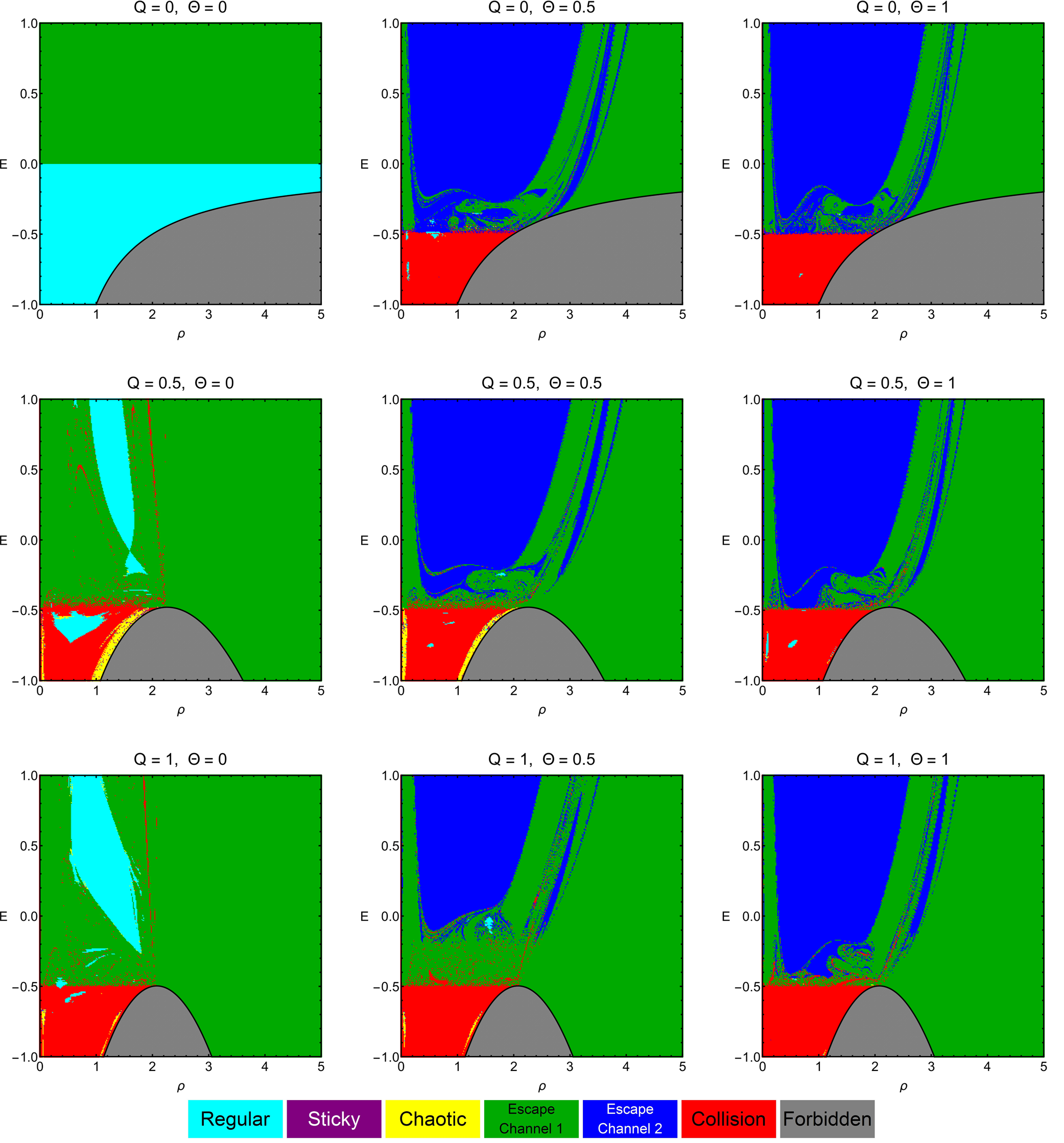}
\caption{Basin diagrams in the $(\rho, E)$-plane. Each panel shows the possible orbits for different values of the quadrupole and octupole moments.}
\label{rE}
\end{figure*}

\begin{figure*}
\centering
\includegraphics[width=\hsize]{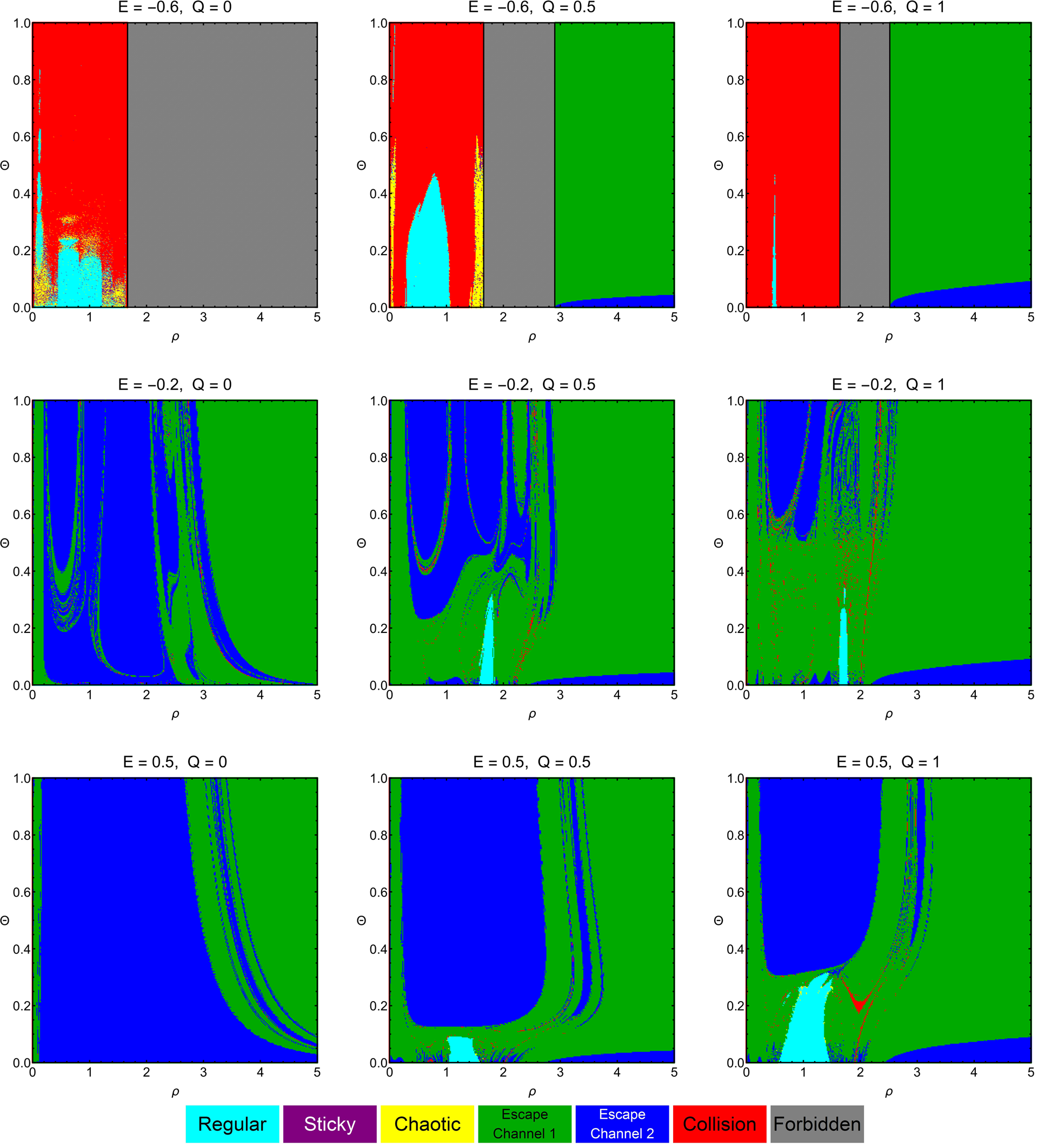}
\caption{Basin diagrams in the $(\rho, \Theta)$-plane. Each panel shows the possible orbits for different values of the energy and quadrupole moment.}
\label{rP}
\end{figure*}

\begin{figure*}
\centering
\includegraphics[width=\hsize]{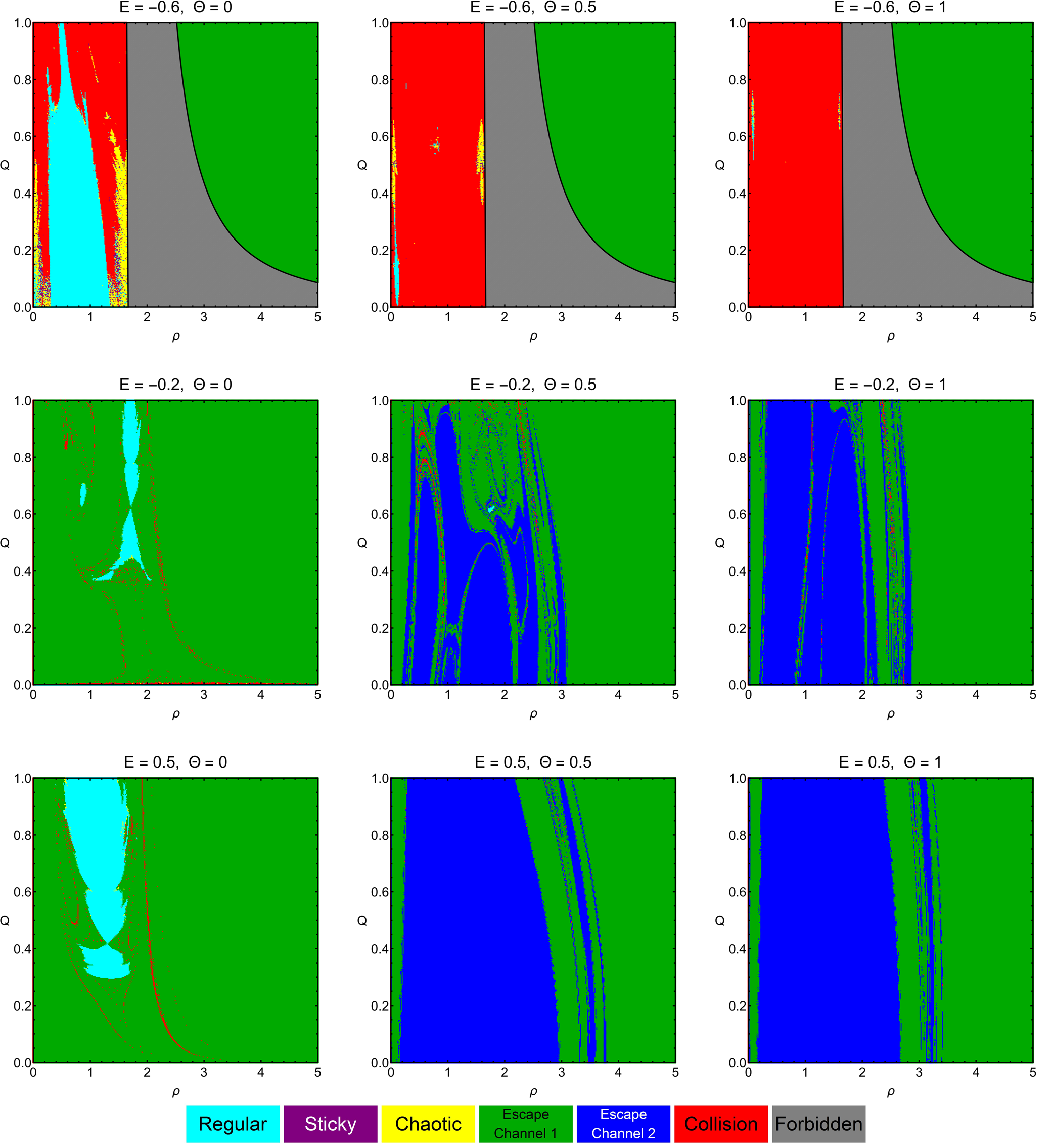}
\caption{Basin diagrams in the $(\rho, Q)$-plane. Each panel shows the possible orbits for different values of the of the energy and octupole moment.}
\label{rQ}
\end{figure*}

\begin{figure*}
\centering
\includegraphics[width=0.7\hsize]{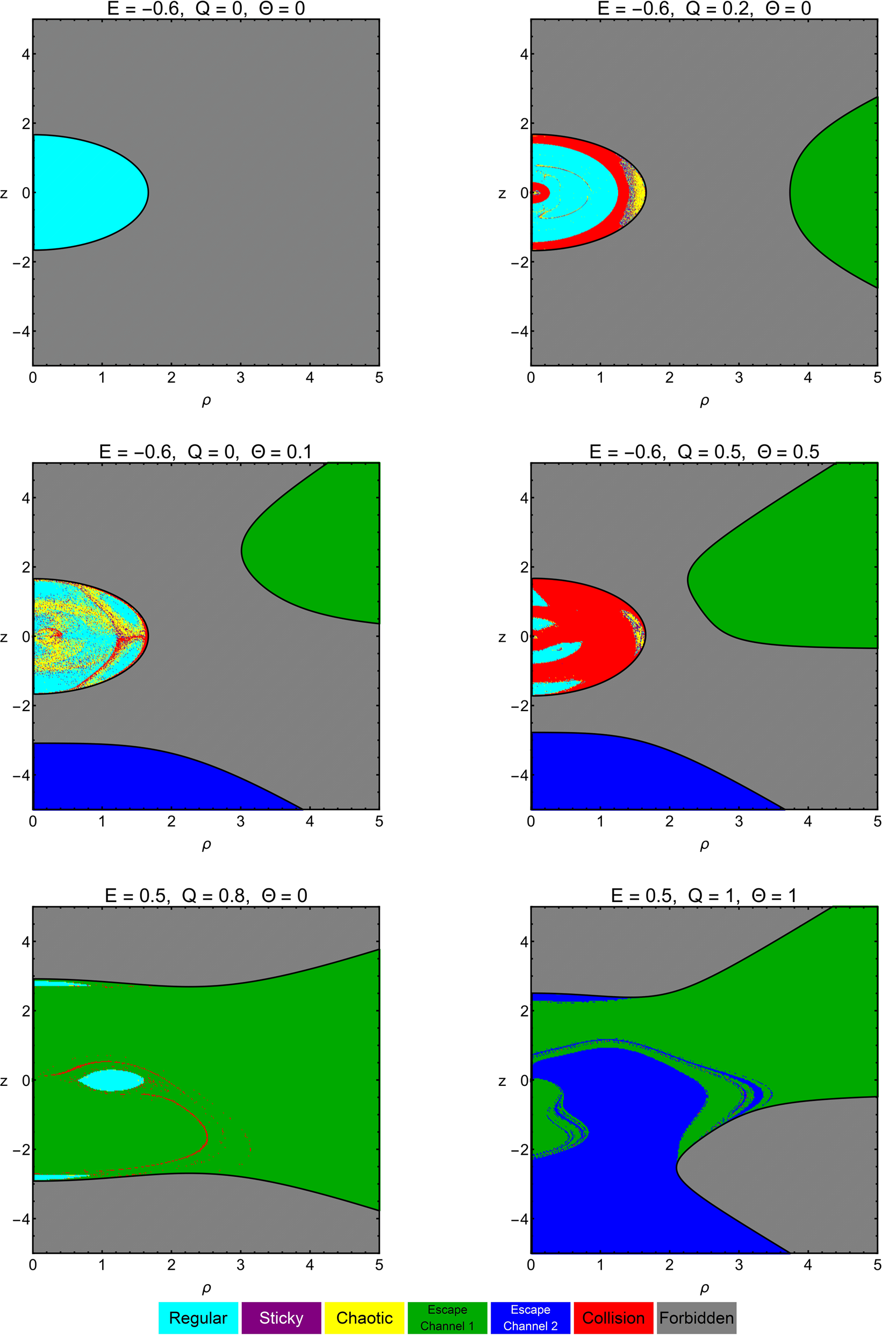}
\caption{Basin diagrams in the $(\rho, z)$-plane. Each panel shows the possible orbits for different values of the of the energy, quadrupole and octupole moments.}
\label{rz}
\end{figure*}

Moreover, the bounded orbits are sub-classified according to its dynamic nature (regular, sticky, or chaotic), while unbounded orbits are sub-classified into escape channels. In Fig.~\ref{orbs} we show an example of each type of orbit present in the current Newtonian system: regular, sticky, chaotic, collision, escape through channel 1, and escape through channel 2.

The method used in this work to distinguish between regular and chaotic orbits is the so-called SALI (for its acronym Smaller Alignment Index) that allows classifying the orbits according to the numerical value obtained after evolving two deviation vectors $\vec{w_{1}}$ and $\vec{w_{2}}$, which must be periodically normalized to avoid overflow \citep{S01,SABV04,BS12}. More specifically, if SALI $>10^{-4}$ the trajectory is classified as regular, while if SALI $<10^{-8}$ it is categorized as chaotic, or if the result belongs to the interval $10^{-4} <$ SALI $< 10^{-8}$, it is classed as sticky and the orbit requires a longer time of integration to be classified. The SALI index is defined as $\mathrm{SALI} \equiv \min \left(d_{-}, d_{+}\right)$, with
\begin{equation}
d_{\mp} \equiv \left\| \frac{\vec{w}_{1}}{\left\|\vec{w}_{1}\right\|} \mp \frac{\vec{w}_{2}}{\left\|\vec{w}_{2}\right\|} \right\|.
\label{sali}
\end{equation}

By using several planes of representation, e.g., $(\rho, z)$, $(\rho, E)$, $(\rho, \Theta)$, and $(\rho, Q)$, in what follows, we have made integrations spanning $10^4$ time units for a fine net of initial conditions inside the scattering regions. The numerical method for integrating the equations of motion is based on a Bulirsch-Stoer algorithm \citep{PTVF92,S18}, in which the numerical errors are of the order $10^{-12}$ (or less). Hereafter, the orbital classification is indicated using color-coded basins diagrams \citep{N04,N05}, with the following color code: (cyan) regular, (purple) sticky, (yellow) chaotic, (red) collisional, (green) escaping through channel 1, and (blue) escaping through channel 2.

In Fig.~\ref{rE} we use different combinations of the multipole moments $(Q, \Theta)$ to illustrate the classification of orbits in an energy versus position graph. In the first case $Q = 0, \Theta = 0$, it is observed that for positive energy values the only possibility is to obtain escape orbits, while if the energy is negative, the system exhibit only regular orbits. This case corresponds to the dynamics of a test particle in presence of a point mass source and coincides with the regular dynamics of the Schwarzschild solution in the GR case.

Displayed on the next two panels of the first row of Fig.~\ref{rE}, the cases of $Q = 0$ and $\Theta \neq 0$ show a very different behavior. For $E > -0.5$, the majority of orbits correspond to escape through the channels 1 and 2, with tiny regions of regular and chaotic motions, while for $E < -0.5$, the graph is filled with collision orbits and small zones of regular motion barely observable to the naked eye. On the other hand, in the next two panels of the first column, the cases $Q \neq 0$ and $\Theta = 0$ are considered. Here, it is observed that escape is only possible through channel 1, and regardless of the value of $E$, for $\rho > 2$ the plane is filled with this type of orbit. When $\rho < 2$ and $E < -0.5$ the collision orbits are the common scenario but with considerable zones of regular and chaotic motions, moreover, when $\rho < 2$ and $E > -0.5$ the set of possible orbits are replaced by escape, collision and wide zones of regular orbits with small traces of chaos. The last four panels of the second and third rows and columns show that exit through channel 2 is possible if and only of $\Theta\neq0$ and that for larger values of the octupole moment $\Theta$ the bounded motion and hence the chaotic and regular zones tend to disappear.

The analysis of this figure suggests an opposite behavior of the dynamics in the Newtonian system in comparison with the relativistic setup, making more evident the need of the quadrupole moment for the presence of chaotic orbits, than the need of a nonzero octupole moment. To clarify this point, in Figs.~\ref{rP} and \ref{rQ} we present the basin diagrams in the $(\rho, \Theta)$ and $(\rho, Q)$ planes respectively, using different values of the energy and multipole moments.

In Fig.~\ref{rP} we present the orbit classification in the $(\rho, \Theta)$ plane for different combinations of  energy $E = -0.6, -0.2, -0.5$ and quadrupole moment $Q = 0, 0.5, 1$. Here, it can be noted that according to the observed in Fig.~\ref{rE}, for energy values larger than $0.5$ (second and third row in Fig.~\ref{rP}) the plane is dominated by escape orbits in which the zones of regular motion appear when $Q \rightarrow 1$. However, when the energy value equals $-0.6$, the system shows zones of collision, regular and chaotic orbits, where the last two types of orbits take place mainly for small values of the octupole moment $\Theta < 0.4$. Also, it is important to note that collision zones completely encompass the scattering region when $Q \rightarrow 1$.

On the other hand, in Fig.~\ref{rQ} we present the orbit classification in the $(\rho, Q)$  plane for different combinations of energy $E = -0.6, -0.2, -0.5$, and quadrupole moment $Q = 0, 0.5, 1$. In this case zones of chaotic and regular motion can be easily observed when the energy takes the value $E = -0.6$, however, these areas shrink for larger values of the octupole moment. Also for $E > -0.5$ (second and third row in Fig.~\ref{rQ}), it is observed that the majority of orbits belong to escape through channels 1 and 2, but contrary to the observed in Fig.~\ref{rP} the zones of regular motion tend to disappear for $Q \rightarrow 1$.

The classification of orbits in the configuration plane $(\rho,z)$ is presented in Fig.~\ref{rz}. From the first two rows of this figure ($E=-0.6$) it can be noted that the regions of allowed motion, as determined by the zero velocity surfaces, can be strongly affected by the multipole moments, however, the new regions that take place when $Q$ or $\Theta$ are greater than zero, do not host bounded orbits and therefore they shall not influence the structure of the astrophysical system. Moreover, it should be noted that the inner semi-elliptical shaped region centered at the origin will change the whole structure of test particles orbiting the source since the perfect spheroid on the top left panel can be completely modified with the appearance of bands of test particles falling into the source. Lastly, in the bottom panels of Fig.~\ref{rz}  we show the case of positive energies ($E = 0.5$). Here two different structures appear, the first one when the octupole moment is zero (bottom left), showing the possibility to have a ring of test particles orbiting the source about $\rho=1$, with two almost flat structures of test particles at the top and bottom of the origin. The second structure (bottom right), corresponds to higher values of the multipole moments where it is observed that all the test particles will be repelled by the source through the escape channels.

\section{Concluding Remarks}
\label{conc}

In the present work, we have studied the orbit classification of test particles in the presence of a Newtonian potential whose relativistic counterpart describes a superimposed halo with a black hole. Taking into account that our system contains as free parameters the energy along with the quadrupole and octupole moments, it is shown this set of parameters define univocally not only the regions of allowed motion but also manage the bounded and unbounded movement. It was found, that the presence of the octupolar moment introduces an asymmetry in the classical system with respect to the $z$-axis. In accordance with the relativistic system for the absence of quadrupole moment, it is possible to get regions of chaotic motion, however, in the absence of the octupolar moment, these chaotic regions can also be found for the Newtonian system. In particular, the multipole moments could certainly modify the structure of test particles orbiting the system with the increase of their values, from a perfectly shaped spheroid when $Q=\Theta=0$ to a set of bands of bounded motion, or even the absence of test particles orbiting the source for $Q=\Theta\rightarrow 1$.

Our results indicate that although chaos and the multipolar moments are intrinsic properties of a given relativistic source, and that the set of relativistic multipole moments of mass reduce to the multipole moments in Newtonian theory, there is no direct correspondence on the classical and relativistic dynamics induced by the multipole moments. This characteristic can be due to: (i) the different conceptions for space and time in both regimes, or (ii) it can be inferred that the only reason that allows the occurrence of chaotic behavior of orbits in relativistic systems is the reflection symmetry breaking about the equatorial plane. To solve these discrepancies, new studies using post-Newtonian approaches could shed light on the reasons for no direct correspondence.

\section*{Acknowledgments}

This project was funded by the Deanship of Scientific Research (DSR) at King Abdulaziz University, Jeddah, Saudi Arabia, under grant no. KEP-17-130-41. The authors, therefore, thankfully acknowledge DSR for technical and financial support. F.L.D was also partially supported by MinCiencias (Colombia) Grant 8863 and by Universidad de los Llanos.


\begin{thebibliography}{00}

\bibitem[\protect\citeauthoryear{Beig \& Simon}{1980}]{BS80} Beig R., Simon, W. 1980, Communications in Mathematical Physics, 78, 75

\bibitem[\protect\citeauthoryear{Bountis \& Skokos}{2012}]{BS12} Bountis T., Skokos Ch. 2012, Complex Hamiltonian Dynamics (Vol. 10). Springer Science \& Business Media.

\bibitem[\protect\citeauthoryear{Ciufolini et al.}{2012}]{CPP12} Ciufolini I., Paolozzi A., Paris C. 2012, Journal of Physics: Conference Series, 354, 012002

\bibitem[\protect\citeauthoryear{Cvitanovic et al.}{2005}]{CAM05} Cvitanovic P., Artuso R., Mainieri R., Tanner G., Vattay G., et al. 2005, Chaos: classical and quantum. ChaosBook. org, Niels Bohr Institute, Copenhagen 69

\bibitem[\protect\citeauthoryear{de Moura \& Letelier}{2000}]{ML00} de Moura A.~P., Letelier P.~S. 2000, Physical Review E, 61, 6506

\bibitem[\protect\citeauthoryear{Drinkwater et al.}{2003}]{DFHMP03} Drinkwater M., Floberghagen R., Haagmans R., Muzi D., Popescu A. 2003, Goce: Esa's first earth explorer core mission Earth gravity field from space-From sensors to earth sciences, Springer-Verlag, New York

\bibitem[\protect\citeauthoryear{Dubeibe et al.}{2007}]{DPS07} Dubeibe F.~L., Pach\'{o}n L.~A., Sanabria-G\'omez J.~D. 2007, Physical Review D, 75, 023008

\bibitem[\protect\citeauthoryear{Fodor et al.}{1989}]{FHP89}  Fodor G., Hoenselaers C., Perj\'es Z. 1989, Journal of Mathematical Physics, 30(10), 2252

\bibitem[\protect\citeauthoryear{Gair \& Mandel}{2008}]{GLM08} Gair J.~R., Li C., Mandel I. 2008, Physical Review D, 77, 024035

\bibitem[\protect\citeauthoryear{Gu\'eron \& Letelier}{2002}]{GL02} Gu\'eron E., Letelier P.~S. 2002, Physical Review E, 66, 046611

\bibitem[\protect\citeauthoryear{Hoenselaers \& Perjes}{1990}]{HP90} Hoenselaers C., Perjes Z. 1990, Classical and Quantum Gravity, 7, 1819

\bibitem[\protect\citeauthoryear{Igata et al.}{2015}]{IIY15} Igata T., Ishihara H., Yoshino H. 2015, Physical Review D, 91, 084042

\bibitem[\protect\citeauthoryear{Laarakkers \& Poisson}{1999}]{LP99} Laarakkers W.~G., Poisson E. 1999, The Astrophysical Journal, 512, 282

\bibitem[\protect\citeauthoryear{Liu et al.}{2017}]{LWH17} Liu L., Wu X., Huang G. 2017, General Relativity and Gravitation, 49(2), 28

\bibitem[\protect\citeauthoryear{Nagler}{2004}]{N04} Nagler J. 2004, Phys. Rev. E, 69, 066218

\bibitem[\protect\citeauthoryear{Nagler}{2005}]{N05} Nagler J. 2005, Phys. Rev. E, 71, 026227

\bibitem[\protect\citeauthoryear{Press et al.}{1992}]{PTVF92} Press H.~P., Teukolsky S.~A., Vetterling W.~T., Flannery B.~P. 1992, Numerical Recipes in FORTRAN 77, 2nd edn. Cambridge University Press, Cambridge

\bibitem[\protect\citeauthoryear{Quevedo}{1990}]{Q90} Quevedo H. 1990, Fortschritte der Physik, 38, 733

\bibitem[\protect\citeauthoryear{Ramos-Caro et al.}{2011}]{RPL11} Ramos-Caro J., Pedraza J.~F., Letelier P.~S. 2011, MNRAS, 414, 3105

\bibitem[\protect\citeauthoryear{Sanabria-G\'omez et al.}{2010}]{SHD10} Sanabria-G\'omez J.~D., Hern\'andez-Pastora J.~L., Dubeibe, F.~L. 2010, Physical Review D, 82(12), 124014

\bibitem[\protect\citeauthoryear{Shampine}{2018}]{S18} Shampine L.~F. 2018, Numerical solution of ordinary differential equations, Routledge

\bibitem[\protect\citeauthoryear{Shibata \& Sasaki}{1998}]{SS98} Shibata M., Sasaki M. 1998, Physical Review D, 58(10), 104011

\bibitem[\protect\citeauthoryear{Skokos}{2001}]{S01} Skokos C. 2001, Journal of Physics A: Mathematical and General, 34, 10029

\bibitem[\protect\citeauthoryear{Skokos et al.}{2004}]{SABV04} Skokos C., Antonopoulos C., Bountis T.~C., Vrahatis M.~N. 2004, Journal of Physics A: Mathematical and General, 37(24), 6269

\bibitem[\protect\citeauthoryear{Sotiriou \& Pappas}{2005}]{SP05} Sotiriou T.~P., Pappas, G. 2005, Journal of Physics: Conference Series (Vol. 8, No. 1, p. 003). IOP Publishing

\bibitem[\protect\citeauthoryear{Vieira \& Letelier}{1996}]{VL96} Vieira M.~W., Letelier P.~S. 1996, Phys. Rev. Lett., 76, 1409

\bibitem[\protect\citeauthoryear{Vieira \& Letelier}{1999}]{VL99} Vieira M.~W., Letelier P.~S. 1999, The Astrophysical Journal, 513, 383

\bibitem[\protect\citeauthoryear{Visser}{1999}]{V99} Visser P.~N.~A.~M. 1999, Advances in Space Research, 23, 771

\bibitem[\protect\citeauthoryear{Wald}{2010}]{W10} Wald R.~M. 2010, General relativity, University of Chicago press

\bibitem[\protect\citeauthoryear{Wang et al.}{2018}]{WCJ18} Wang M., Chen S., Jing J. 2018, Physical Review D, 98(10), 104040

\end{thebibliography}
\end{document}